\documentclass[useAMS,usenatbib]{mn2e}
\usepackage{amsmath}
\usepackage[dvips]{graphicx}

\usepackage{graphicx, epsfig}
\usepackage{graphics}
\newcommand{\galics}{{\sc galics}}
\newcommand{\momaf}{{\sc momaf}}
\newcommand{\lemomaf}{{\sc lemomaf}}

\newcommand{\beq}{\begin{eqnarray}}
\newcommand{\eeq}{\end{eqnarray}}
\newcommand{\apj}{ApJ}

\newcommand{\apjl}{ApJL}
\newcommand{\aj}{AJ}
\newcommand{\mnras}{MNRAS}
\newcommand{\aap}{A\&A}

\newcommand{\nat}{Nature}

\newcommand{\prd}{PhysRevD}
\newcommand{\aph}{APh}
\newcommand{\avg}[1]{\langle{#1}\rangle}
\title[\lemomaf{}]
      {LeMoMaF: Lensed Mock Map Facility}
\author[J.E. Forero-Romero et al.]
       {Jaime E. Forero-Romero$^{1}$\thanks{E-mail:forero@obs.univ-lyon1.fr(JEFR)},
         J\'er\'emy Blaizot$^{2}$,
        Julien Devriendt$^{1}$,
        \and Ludovic Van Waerbeke$^{3}$,
        Bruno Guiderdoni$^{1}$ \\
        $^{1}$Universit\'e Claude Bernard Lyon 1, CNRS UMR 5574, ENS Lyon,
        Centre de Recherche Astronomique de Lyon, \\
        $\ \ $Observatoire de Lyon, 9 Avenue Charles Andr\'e, 69561 St-Genis-Laval Cedex, France\\
        $^{2}$Max-Planck-Institut f\"ur Astrophysik, Karl-Schwarzschild-Str. 1, 85741 Garching, Germany\\
        $^{3}$University of British Columbia, 6224 Agricultural Road, Vancouver, V6T1Z1,B.C.,Canada\\
      }
\begin{document}

\date{Accepted. Received ; in original form}

\pagerange{\pageref{firstpage}--\pageref{lastpage}}

\maketitle
\label{firstpage}

\begin{abstract}
We present the {\it Lensed Mock Map Facility} (LeMoMaF), a tool
designed to perform mock weak lensing measurements on numerically
simulated chunks of the universe. Coupling N-body simulations to a
semi-analytical model of galaxy formation, LeMoMaF can create
realistic lensed images and mock catalogues of galaxies, at
wavelengths ranging from the UV to the submm. To demonstrate the
power of such a tool we compute predictions of the source-lens
clustering effect on the convergence statistics, and quantify the
impact of weak lensing on galaxy counts in two different filters. We
find that the source-lens clustering effect skews the probability
density function of the convergence towards low values, with an
intensity which strongly depends on the redshift distribution of
galaxies. On the other hand, the degree of enhancement or depletion
in galaxy counts due to weak lensing is independent of the
source-lens clustering effect. We discuss the impact on the
two-points shear statistics to be measured by future missions like
SNAP and LSST. The source-lens clustering effect would bias the
estimation of $\sigma_8$ from two point statistics by $2\% -5\%$. We
conclude that accurate photometric redshifts for individual galaxies
are necessary in order to quantify and isolate the source-lens
clustering effect.
\end{abstract}

\section{Introduction}

One of the key goals of modern astrophysical and astronomical research is to map
out the spatial distribution of the various matter components of the Universe.
In the actual concordance cosmological model, about 85\% of the matter in the universe is
thought to be non-baryonic and non-interacting dark
matter, the other 15\% being composed of baryons \citep{WMAP3}.

One of the most promising tools to track the distribution of dark matter on cosmological
scales is weak gravitational lensing, which has already proven to be well suited
for other purposes, such as precision cosmology \citep{Hu, Huterer,
Benabed, Bernstein, Ismael}.

Weak gravitational lensing affects the observed galaxy properties
such as ellipticities, magnitudes and apparent positions in the sky.
In the weak lensing regime, these effects can only be measured in a
statistical sense. Indeed, the first detections of the weak lensing
signal \citep{B2000, KWL2000, MVWM,
  VW2000, VW2002, Brown03, Jarvis03}, are based on an analysis of the spatial correlation between
ellipticities of galaxies. The study of weak lensing through the changes
in magnitudes and apparent position in the sky that is causes is an even more challenging
measurement \citep{Scranton05, Zhang}.

Most of the difficulties in measuring the weak lensing signal come
from observational systematics such as uncertainties in the
determination of the point-spread function (PSF) and selection
biases \citep{Kaiser, ErbenEtal01, ValeEtal}. There also exist
astrophysical errors related to uncertainties in photometric
redshift calibration \citep{IH05,VWcalibration}, intrinsic
alignments due to physically associated lens-source pairs
\citep{Mandelbaum} or even distant lens-source pairs \cite{HS04}.
Most of the methods used to assess the importance of all these
effects are theoretical or numerical. For the latter, one should also
account, in principle, for uncertainties associated with numerical
errors inherent to N-body simulations, neglected baryonic
cooling effects and to a larger extent a poor understanding
of the theory of galaxy formation.

The most realistic simulations of weak gravitational lensing
performed so far rely on a dark matter distribution taken from an
N-body simulation and galaxies drawn from a Halo Occupation Model
(HOD) parameterized by halo mass alone \citep{H06,VWcalibration},
without any information about apparent magnitudes or galaxy colors.
Even if this information was available from such models, the mere
fact that properties of dark matter halos depend not only on their
masses, but also on their assembly histories --- and that these
latter have significant effects on galaxy clustering \citep{Croton}
--- rules out HOD models based on halo mass as precision tools to
model effects that rely heavily on a realistic description of the
correlations between galaxy and dark matter distributions. The
consensus is that an analysis which includes more realistic galaxy
populations at various wavelengths are needed.

This paper presents an approach which is a first step to fill this gap and
perform more realistic weak lensing simulations thanks to a more realistic
description of galaxy populations. It consists in building a tool which we call
Lensed Mock Map Facility (\lemomaf{}) hereafter, to link together three well
established numerical techniques, each one tackling a different aspect of the problem.
More specifically, a state-of-the-art semi-analytic model (SAM) tracks the properties
of galaxies within a high resolution N-body simulation as they evolve in time. Then a
light cone is assembled from the outputs of the simulations
to convert ``theoretical'' quantities into observables. Finally,
a ray tracing algorithm extracts the weak lensing signal from the cone.
The SAM of galaxy formation that we use in this work is \galics{} (GALaxies In Cosmological
Simulations, \citep{HattonEtal03}), and mock galaxy maps along with dark matter cones are
obtained through a random tiling technique of simulation snapshots, using the
\momaf{} (Mock Map Facility, \citep{momaf}) pipeline.

This paper is organised as follows. In Sec. \ref{sec:model} we review
the characteristics of \galics{} and \momaf{} which are relevant to
the present study. In Sec. \ref{sec:lemomaf} we explain the weak
lensing formalism and its implementation in \lemomaf{}. In
Sec. \ref{sec:results} we present the statistics of the weak lensing
convergence and the effect of weak lensing on differential galaxy
counts. Finally, we discuss our results and outline prospectives in Sec. \ref{sec:conclusion}

\section{GalICS and MoMaF}\label{sec:model}
\subsection{\galics{}}

\begin{figure*}
\begin{center}
\includegraphics[scale=0.40]{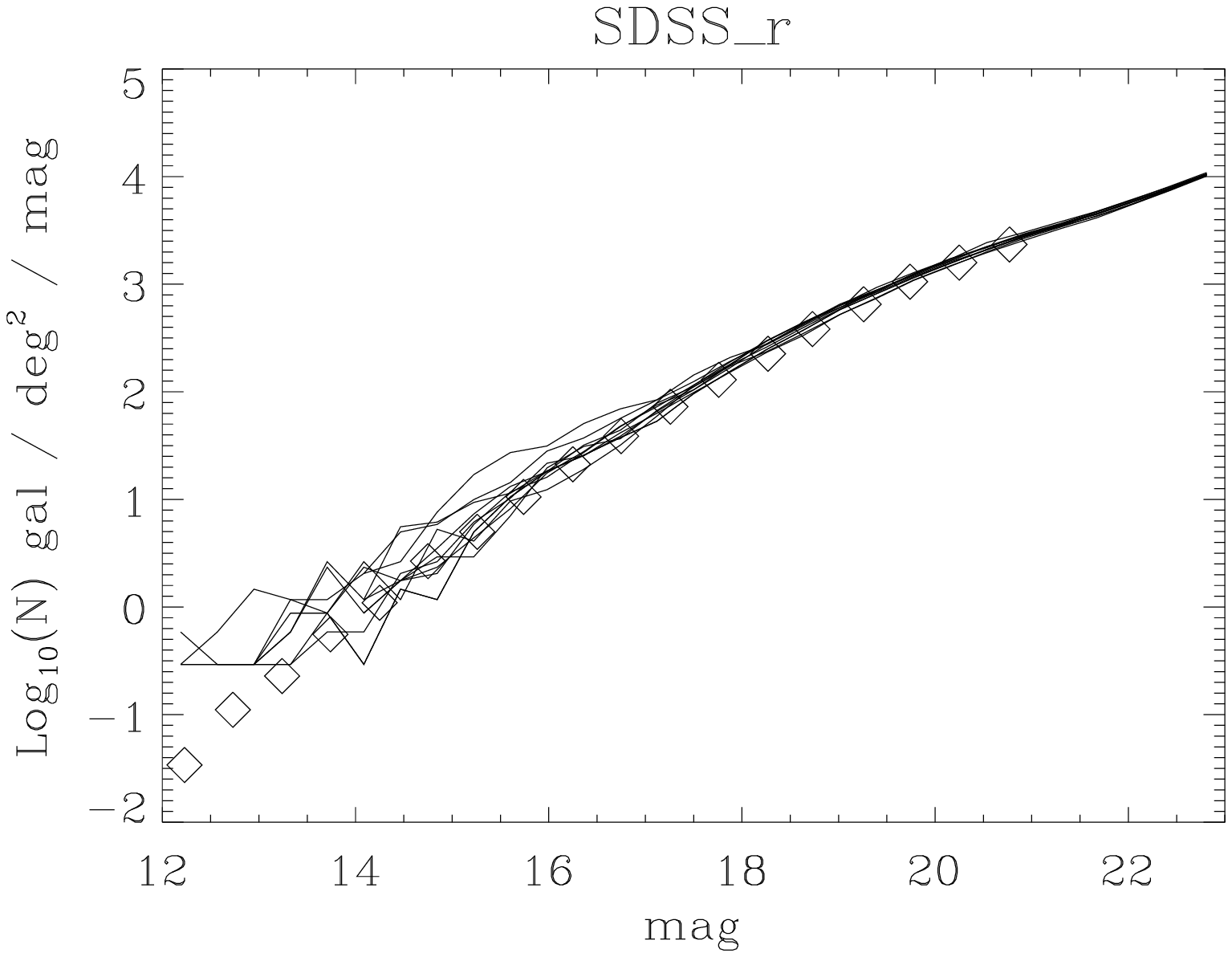}
\includegraphics[scale=0.40]{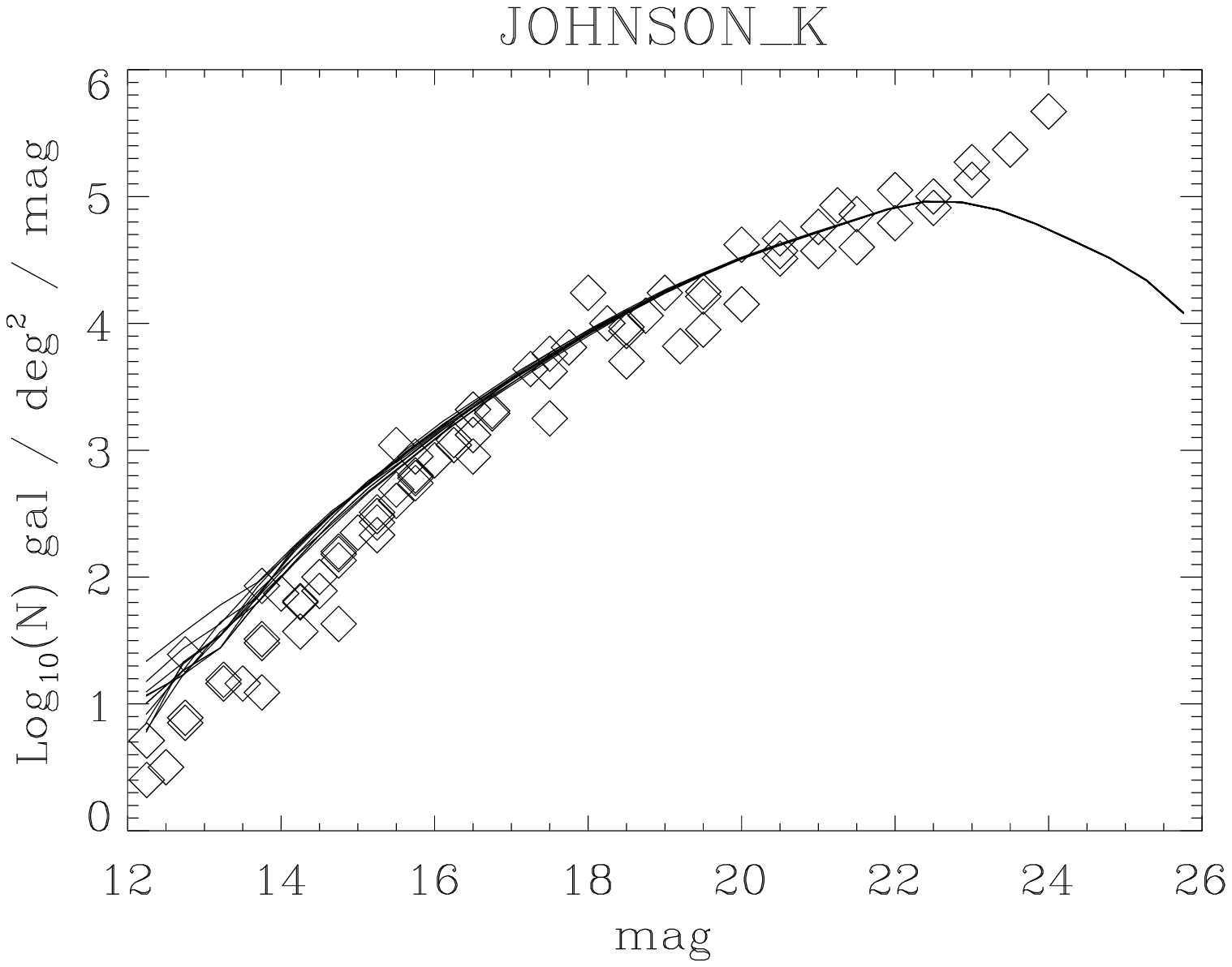}
\end{center}
\caption{\label{gcounts} Galaxy counts in the filters {\tt SDSS r}
  \citep{YasudaEtal01} and {\tt JOHNSON K} \citep{Djorgovski, Gardner,
    Moustakas} obtained through the \galics{}-\momaf{}
  pipeline. Squares are the observational points, and lines the
  results of the analysis of several virtual light cones built with
  the simulation.
  On the right hand side panel one can clearly see when the incompleteness in the
number of low luminosity mock galaxies kicks in due to the finite mass resolution of the simulations.}
\end{figure*}

\galics{} is a hybrid model of galaxy formation which combines
cosmological dark matter N-body simulations with a semi-analytic description of
baryonic processes. The model is fully described in
\citet{HattonEtal03}, and the version we use here is the same as that
used in the previous papers of the \galics{} series
\citep{HattonEtal03, DevriendtEtal05, BlaizotEtal04}. We briefly
recall what are the main ingredients in Appendix \ref{sec:dm} and
\ref{sec:baryons}. Eventually, \galics{} outputs are turned into mock
catalogues using \momaf{}. The following section summarizes how this is achieved.

\subsection{\momaf{}}\label{sec:momaf}

\begin{figure*}
\begin{center}
\includegraphics[scale=0.55]{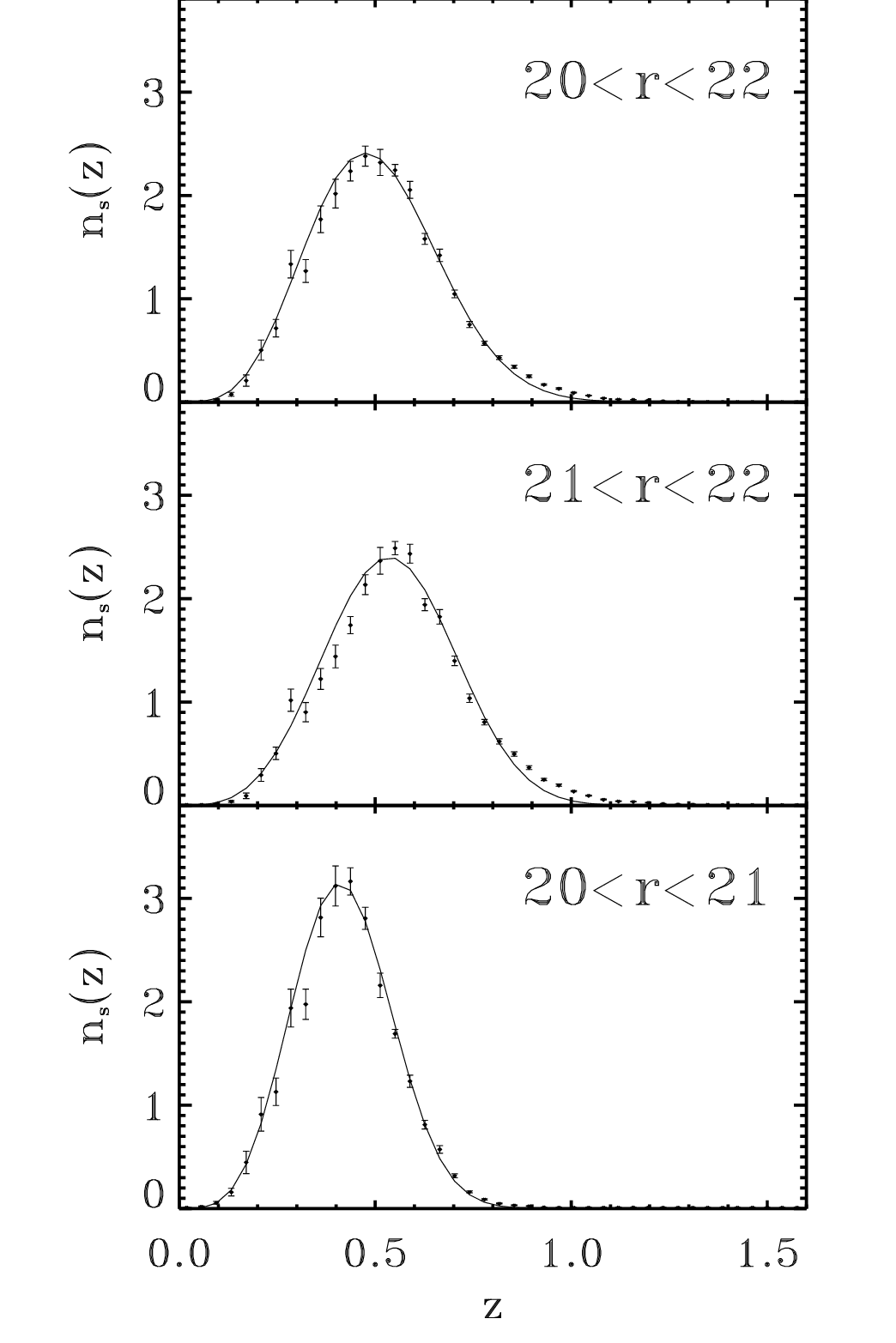}\hspace{1.5cm}
\includegraphics[scale=0.55]{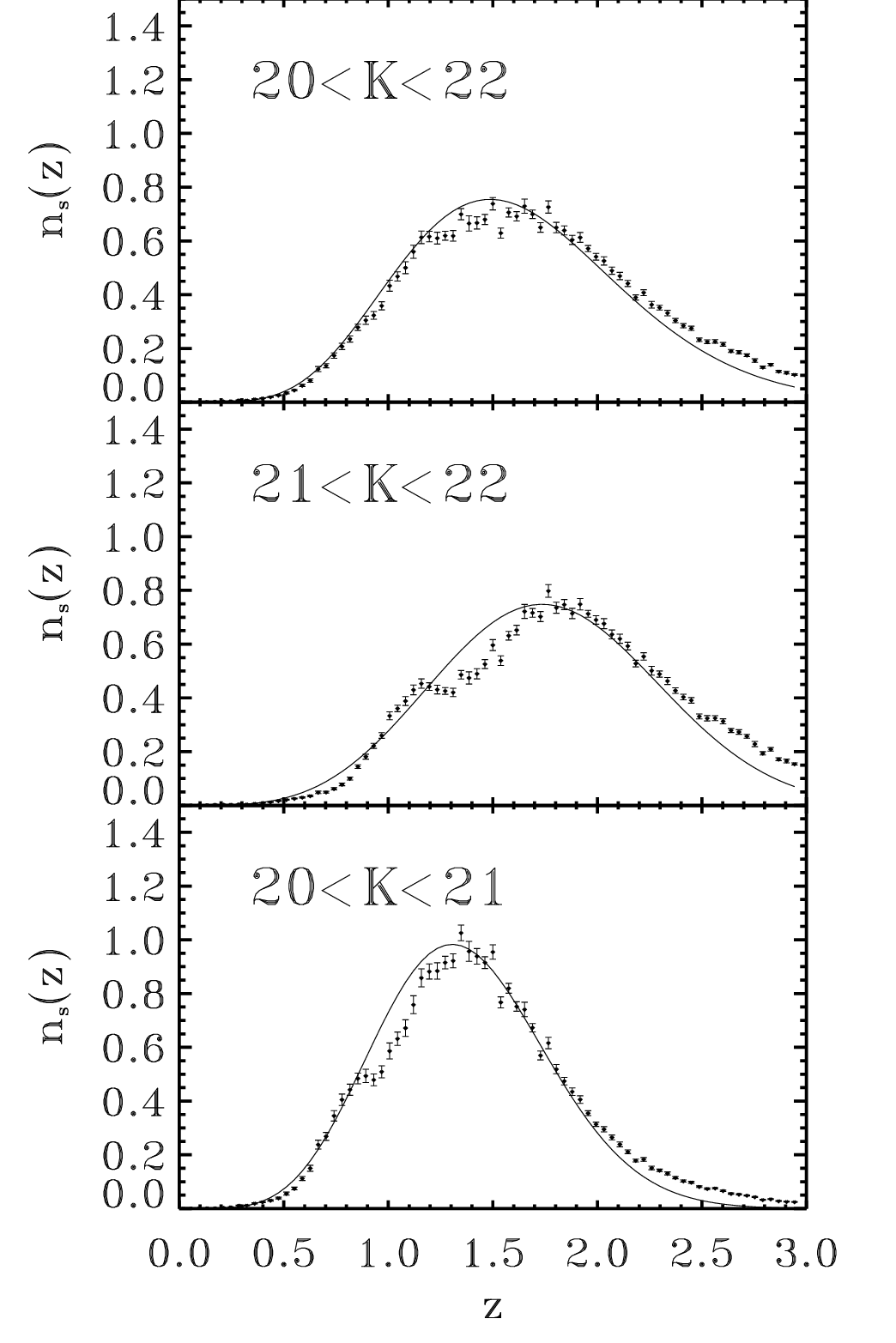}
\end{center}
\caption{ \label{distros} Distributions in redshift for different cuts
in $r$ and $K$ magnitude of the galaxy cones. The dots show the
mean value of $n_{s}(z)$ obtained by averaging over the 25 cones, and the error
bars show the 1-$\sigma$ error on this mean value. The continuous line
is the fit to the function in Eq. \ref{fit}.}
\end{figure*}

\momaf{} \citep{momaf} is a tool which converts theoretical outputs of
hierarchical galaxy formation models into catalogues of virtual
observations. The general principle is simple: mock light cones
are generated using semi-analytically post-processed snapshots of
cosmological N-body simulations. These cones can then be projected to
synthesise mock sky images.

\momaf{} uses a {\it random tiling} technique described in
\cite{momaf} to build mock observations from the redshift outputs of
\galics{}. As explained in \cite{momaf}, several different observing
cones can be generated from the same set of outputs of \galics{}, by
changing either the line of sight or the seed for the random tiling.
\cite{momaf} also discuss the bias induced by this random tiling
approach on the clustering signal of the final maps.

In this paper, we build 25 cones of galaxies and 25 corresponding
cones of dark matter with seeds and
lines-of-sight randomly chosen for each observer position, using the
same seed for every matching pair of galaxy and dark matter cone. These 25
cones allow us to infer an estimate of the dispersion  of clustering
measurements, that is to say, an estimate of the cosmic variance associated with our mock
catalogues.  We use galaxy magnitudes in two filters: {\tt SDSS r}
and {\tt JOHNSON K}, because the results in these bands have already
been thoroughly examined by \cite{momaf} and \cite{galicsV}. The
results of the galaxy counts from \galics{}-\momaf{} and its
comparison to observations are shown in Fig. \ref{gcounts}, and galaxy redshift
distributions for different cuts in magnitude are shown in
Fig. \ref{distros}. In order to make semi-analytic predictions of the weak lensing
statistics \citep{VW01} for this broad distribution we describe it with the following functional form:
\begin{equation}
\label{fit}
n(z) = \frac{\beta}{z_{0}\Gamma\left(\frac{1 + \alpha}{\beta}\right)}\left(\frac{z}{z_{0}}\right)^{\alpha}\exp\left[-\left(\frac{z}{z_0}\right)^{\beta}\right]
\end{equation}
where $\Gamma(x)$ is the Gamma function and $\alpha$, $\beta$ and $z_{0}$ are free parameters to determine for each distribution.

However, we note that given the rather small size of the original N-body simulation
box ($100$ $h^{-1}$ Mpc on a side) that was used to run \galics{} on,
our estimate of the cosmic variance is likely to be biased
and has to be taken as a lower boundary on the true cosmic errors. Finally, the mock
catalogues we will use in the following are $4400$ $h^{-1}$ Mpc in depth, and
$1.4^{\circ}\times 1.4^{\circ}$ in angular size.

\section{Lensed MoMaF}\label{sec:lemomaf}
\subsection{Weak Lensing Formalism}
In this section we provide a summary of the weak lensing equations relevant for
building \lemomaf{} and refer the reader interested in a more detailed treatment to
 \cite{JainEtal00}.

We use comoving coordinates and place ourselves in the framework of an
isotropic and homogeneous universe that can be described  by the
Robertson-Walker metric, where in the presence of a perturbative
gravitational potential the change in a photon's direction can be
written as:

\begin{equation}
d\vec{\alpha} = -2\vec{\nabla}_{\perp}\phi d\chi
\end{equation}

where $d\vec{\alpha}$ is the photon's deviation, $\phi$ is the gravitational
potential, $\vec{\nabla}_{\perp}$ is the gradient in the direction
perpendicular to the photon line of propagation and $\chi$ is the radial
comoving coordinate. A deflection at $\chi^{\prime}$ produces in a plane
located at coordinate $\chi$, perpendicular to the line of sight, a deflection
of

\begin{equation}
d\vec{x}= r(\chi - \chi^{\prime})d\vec{\alpha}(\chi^{\prime})
\end{equation}
where $r(\chi)$ is the comoving angular distance, that in a plane universe
($\Omega_{\Lambda}+\Omega_{m}=1$) is equal to $\chi$.
Integrating along the perturbed trajectory of the photon, and then dividing by
$r(\chi)$ we obtain the angular position at position $\chi$

\begin{equation}
\vec{\theta}(\chi)=\vec{\theta}(0) - \frac{2}{r(\chi)}\int_{0}^{\chi}d\chi^{\prime}r(\chi-\chi^{\prime})\vec{\nabla_{\perp}}\phi
\end{equation}

This treatment holds for a single photon. To obtain the effect on an extended object we
calculate the Jacobian of the transformation
($\partial\theta_{i}(\chi)/\partial\theta_{j}(0)$), which we call $A_{ij}$ where
$i$ and $j$ denote perpendicular directions in the plane perpendicular to the line of sight,
in which $\vec{\theta}$ is also located:

\begin{equation}
A_{ij} = -2\int_{0}^{\chi}d\chi^{\prime}\frac{r(\chi - \chi^{\prime}) r(\chi^{\prime})}{r(\chi)} \nabla_{i}\nabla_{j}\phi + \delta_{ij}
\end{equation}
the Jacobian matrix is usually decomposed as follows:
\begin{equation}
\mathbf{A}=\begin{pmatrix}
  1 - \kappa - \gamma_{1} & -\gamma_{2} - \omega\\
  -\gamma_{2}+\omega & 1 -\kappa + \gamma_{1}
\end{pmatrix}
\end{equation}
where $\kappa$  is the convergence, $\gamma$ is the shear and $\omega$ the
rotation.

The relation between the gravitational potential and the mass density
perturbation  $\delta = \rho/\bar{\rho} - 1$ is given by:
\begin{equation}
\nabla^{2}\phi =
\frac{3}{2}\left(\frac{H_{0}}{c}\right)^{2}\Omega_{m}\frac{\delta}{a}
\label{poisson}
\end{equation}
where $H_{0}$ is the Hubble constant at the present, $c$ is the speed of light
and $a$ is the expansion factor.

\subsection{Multiple plane formalism}
The multiple plane formalism consists in dividing the space between the source and the
observers into $N$ equidistant (in comoving coordinates) planes perpendicular to
the line of sight.
Following the convention of \cite{HM} we place ourselves in a cartesian coordinate
system noted by ($x_1, x_2, y$) where the origin is the observer and $y$
indicates the direction of observation. For small angular size fields of view,
$y$ can be identified with the radial comoving distance
$\chi$. The inter-plane comoving distance will be called $\Delta y$.
The projected density contrast on the $i$-th plane located at $y_{i} $ is
given by:

\begin{equation}
\Sigma_{i}(x_1,x_2) = \int_{y_{i-1}}^{y_{i}}dy\delta(x_1,x_2,y)
\end{equation}
which defines a corresponding two dimensional potential $\Psi = 2 \int dy \phi$.
The position of a light ray in the $n$-th plane can then be found using the multiple
plan lens equation:

\begin{equation}
\vec{\theta}_{n} = \vec{\theta}_{1} -\sum_{i=1}^{n-1}\frac{r(\chi_{n}
  - \chi_{i})}{r(\chi_{n})}\vec{\nabla}\Psi_{i}
\label{propaga}
\end{equation}
\begin{equation}
  \mathbf{A}_{n} = \mathbf{I} - \sum_{i=1}^{n-1}\frac{r(\chi_{n} -
    \chi_{i})r(\chi_{i})}{r(\chi_{n})}\mathbf{U}_{i}\mathbf{A}_{i}
\label{eq:amplification}
\end{equation}
where $\mathbf{I}$ is the identity matrix and $\mathbf{U}_{i}$ is defined by:

\begin{equation}
\mathbf{U}_{i} =\begin{pmatrix}
\partial_{11}\Psi_{i} & \partial_{12}\Psi_{i}\\
\partial_{21}\Psi_{i} & \partial_{22}\Psi_{i}
\label{eq:U}
\end{pmatrix}
\end{equation}
where $\partial_{ij}$ symbols stand for partial differentiation.

\subsection{Ray Tracing Algorithm}
\label{programa}
The algorithm can be divided in three steps. First, the projection of the dark matter
distribution onto the planes. Second, the calculation of the potential and its first and
second derivatives in these planes and finally the ray tracing from
the observer to the last plane. In practice, things proceed as follows:

1. Once we have the dark matter particles of the simulation positioned inside an
   light cone build with \momaf{}, we
   define an orthogonal coordinate system ($x_1, x_2, y$) at the origin of the
   cone, where the $y$ axis is directed from the observer along the symmetry
   axis of the cone. We then project (along the $y$ axis) all the particles onto
   $N$ equidistant planes parallel to the one located at the origin of the cone,
   which means for example that particles originally between ($x_1$, $x_2$, $y$) and
   ($x_1$,$x_2$,$y + \Delta y$) now sit at ($x_1$,$x_2$,$y + \Delta y$).

2. In each plane we interpolate the
   overdensity ($\delta$ in equation (\ref{poisson})), using a cloud-in-cell (CIC) algorithm, over an uniform square
   grid, $G$, of $N_g$ cells on a side of comoving size $L_{side}$. $N_{g}$ and
   $L_{side}$ are kept identical for all planes.
   In order to solve the Poisson equation we
   pad with zeroes around the grid $G$, thus creating a grid of size
   $2\times N_g$ in side. We then perform a fast Fourier transform (FFT) of
   the overdensity grid, and in Fourier space we divide by
   $-1/k^2$ to inverse transform and obtain the gravitational
   potential $\Psi$ on the grid. Finally, using finite differences methods
   we obtain the first and second derivatives of the potential along
   the $x_1$ and $x_2$ directions. This allows us to construct the
   matrix $\mathbf{U}$ in Eq. \ref{eq:U} for each point in the plane. A detailed
   numerical implementation for this step can be found in
   \cite{premadi}.

3. We initialize the ray positions over a uniform grid on the nearest
   plane to the observer. For each ray we interpolate the values of
   $\vec{\alpha}$ and the elements of $\mathbf{U}$ using their
   tabulated values on the CIC grid. Applying
   equation (\ref{propaga}) we then figure out the position of each ray on the next plane,
    and again interpolate the values of $\vec{\alpha}$ and $\mathbf{U}$ at this new position
   to calculate $\mathbf{A}$, and so on and so forth until we reach the final plane.
   We store all these values, for each ray on every plane.

\subsection{Shearing of galaxies}

The galaxies in \galics{} are represented by three components: disk, bulge and
burst. Geometrically speaking, on a mock image, the disk is seen as an ellipse, and the bulge
and the burst as circles. The burst is treated as a punctual source and only its
magnitude will be modified by weak lensing. The circle and the ellipse can both be
parametrised by their shape matrix, according to their weighted quadrupole moments.

For an ellipse of major axis $a$, minor axis $b$ and with the major
axis making an angle $\beta$ with respect to an horizontal reference
axis, the shape matrix can be written as:

\begin{equation}
\mathbf{Q} =\frac{1}{\pi ab}\begin{pmatrix}
 a^2\cos^{2}\beta + b^2\sin^{2}\beta & (a^2 -  b^2) \sin\beta\cos\beta\\
 (a^2 -  b^2) \sin\beta\cos\beta       & a^2\sin^{2}\beta + b^2\cos^{2}\beta&
\end{pmatrix}
\end{equation}

The case of the circle is recovered by setting  $a=b$ and $\beta = 0$.
Supposing that the Jacobian matrix does not vary much along the
corresponding surface of the galaxy (which is a good approximation in the
case of weak lensing caused by large scale structure) the lensed
shape matrix $\mathbf{Q^{\prime}}$ can be written as:

\begin{equation}
\mathbf{Q^{\prime}} = \mathbf{A}^{-1}\mathbf{Q}\mathbf{A}^{-1}
\end{equation}

In this way the lensed properties of the disk and the bulge can be
easily found if one knows the Jacobian matrix. In \lemomaf{} the $\mathbf{A}$ matrix
used to lens a galaxy is the average of four matrices, located at the
point of impact of the four closest rays on the nearest plane to this galaxy. We lens bulge and disks
separately.

This is used to produce mock lensed images useful for a more realistic
treatment of the detection in simulations of the lensing signal through cosmic
shear. In this paper we won't use this capability. We will infer the
measurements of $\kappa$ directly from the numerical values used to
modify the galaxy's properties.

\subsection{Limits of the method}
\label{sec:limits}

\begin{figure}
\begin{center}
\includegraphics[scale=0.35]{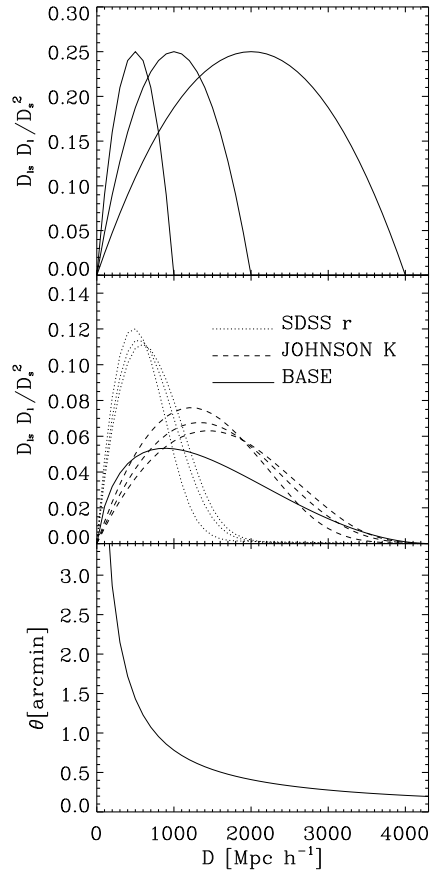}
\end{center}
\caption{\label{resolution} {\it Top Panel}: Distance combination
in equation \ref{eq:amplification} normalized to the distance to the
source. The three curves correspond to a distance to the source of
$1000$, $2000$ and $3000$ $h^{-1}$ Mpc. {\it Middle panel}: Sum over
the curves of top panel weighted by $n_s(z)$ at the redshift of the source
plane. The solid curve, labeled as {\tt BASE}, shows this sum for a
population of galaxies homogeneously distributed in redshift. The curves labeled as
{\tt JOHNSON K} and {\tt SDSS r} show the sum for the respective
galaxy distribution in Fig.\ref{distros}. {\it Lower panel}: value of
the angular resolution in our ray tracing simulation as a function of
distance from the observer.}
\end{figure}

Most of the limitations of our ray tracing implementation are resolution issues that come
from the use of N-body simulations and the interpolation grids used to
trace the light rays. These errors in the calculation of the
weak lensing signal from numerical simulations have been thoroughly
quantified before \citep{JainEtal00, vw}. We recall these results here,
explicitly pointing out the place where the parameters that we use
limit the approach the most.

Two relevant scales can be defined to asses the quality of
the weak lensing signal in the simulation. The first one,
$\sigma_{g}$, relates to the size of the Fourier grid where the
lensing signal is obtained, and the second, $\sigma_{n}$, to the
finite resolution of the N-body simulation. We define these quantities
as: $\sigma_{g}= L_{side}/N_{g}$, $\sigma_{n} =
L_{box}/N_{part}^{1/3}$, where $L_{side}$ is the size of the grid in the
plane where we interpolate, $N_{g}$ its number of cells in one dimension, $L_{box}$ is the size of the
simulation box and $N_{part}$ is the number of particles inside this box.

If $\sigma_{g}$ is larger than $\sigma_{n}$, the power of the signal
on small scales will be dominated by the Fourier grid cut-off. If it is smaller,
the power of the signal is dominated by noise in the N-body simulation.
Moreover, if $\sigma_{g}$ is not only larger than $\sigma_{n}$ but also larger than
any significant physical scale in the simulation,  the
features of the overdensity field that produce the deflection of the
rays on that scale are wiped out, and no weak lensing signal is measured by the ray tracing method.

We choose a value of $\sigma_{g}/\sigma_{n}\sim 1$, guided by the
results of previous studies. In our case, since we use a simulation with
$L_{box} = 100$ $h^{-1}$ Mpc and $N_{p} = 256^3$, this translates into a
resolution of $\sigma_{n} = 390$ $h^{-1}$ kpc. Our ray tracing
simulation uses a grid with $L_{side}=200$ $h^{-1}$ Mpc and $N_{g}=800$, which translates
into $\sigma_{g}=250$ $h^{-1} kpc$.

For a source at a given redshift, the weak lensing signal probes
structures at intermediate redshifts between source and observer. This can be seen from the
equation (\ref{eq:amplification}) where the distance combination $r(\chi_{n} -
\chi_{i})r(\chi_{i})/r(\chi_{n})$ plays the role of an efficiency
function for the lensing convergence. This distance combination
peaks at an intermediate redshift between the observer and the source as
shown in the upper panel of Fig. \ref{resolution} where it
is plotted normalized to the plane distance for three different source positions.

When one takes into account the redshift distribution of the sources used to measure the weak
lensing signal, $n_s(z)$, one realizes that each redshift contributes a different amount to
each plane. In order to estimate the joint contribution from
the lens efficiency and the redshift distribution of the sources,  we
consider  that each efficiency curve is multiplied by the value of
$n_s(z_{i})$ where $z_{i}$ is the redshift at which the efficiency curve is
calculated, and then we add all the different efficiency curves together. In the middle panel of
Fig. \ref{resolution} we show the results of this calculation for a uniform $n_s(z)$ as well as for
the six source distributions presented in Fig. \ref{distros}.

From the middle panel of Fig. \ref{resolution}, one can see that the distances probed by galaxy
populations measured in the {\tt SDSS r} filter are comprised between $0-1000$
$h^{-1}$ Mpc (redshifts $0$ to $0.5$), while for the {\tt JOHNSON K}
filter this distance interval lies between $1000 - 3000$ $h^{-1} Mpc$
(redshifts from $0.5$ to $2.0$), with a broad peak around $1500$ $h^{-1} Mpc$.

Once the value of $\sigma_{g}$ is fixed, we can determine the
angular resolution as a function of the distance from the observer, as shown
in the lower panel of Fig. \ref{resolution} for our chosen value of $\sigma_{g}$.
From this figure, one can check that in the redshift range probed by
galaxies seen in the {\tt SDSS r} filter this angular resolution is on the order of $\sim\
1.5$ arcmin, and for that probed by galaxies in the {\tt JOHNSON K} filter of about $\sim\ 0.5 arcmin$.

Discontinuities induced by the random rotations and origin shifts of the
tiled boxes in the construction of the cone, using planes that are not
perpendicular to the path of the light rays produce artifacts
that were also analyzed by \cite{vw}, concluding that their effects on
the lensing signal are negligible.
Finally, other errors arise from the translation of dark halo mass resolution
into completeness limits at a given magnitude in a given waveband at a given redshift
for the galaxy population, or from reduced spatial correlations caused by the random tiling
technique. All this limitations have been studied in detail
\cite{HattonEtal03}, \cite{BlaizotEtal04}, \cite{momaf} and \cite{galicsV}. In this
paper we remain inside these limits to draw our conclusions, but explicitly indicate
when they are reached.

\section{Results}\label{sec:results}

In this study, we make two kinds of numerical experiments with
\lemomaf{}: (i) we measure the weak lensing signal induced by the dark matter cone
at galaxy positions, and (ii) we study the change in galaxy counts
caused by weak lensing effects. Before exploring the results of these
experiments we briefly discuss the behaviour of the \lemomaf{} ray tracing module.

\subsection{Ray Tracing Results}

\begin{figure}
\begin{center}
\includegraphics[scale=0.30]{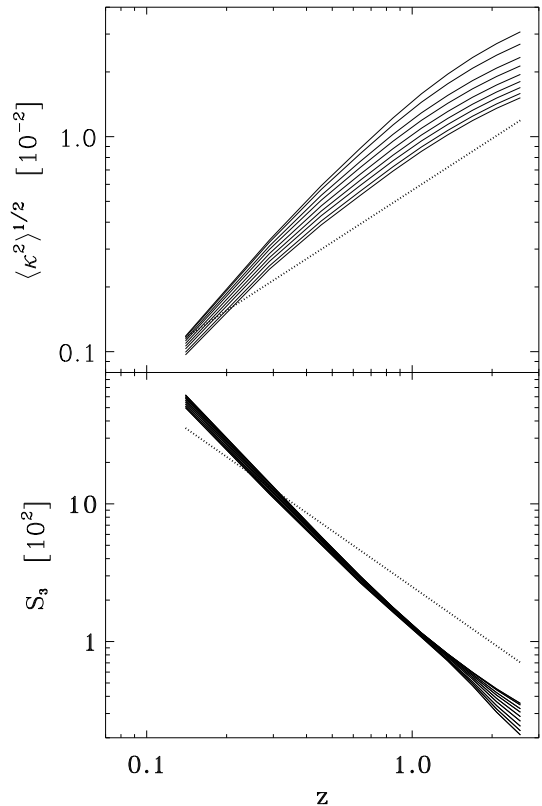}
\end{center}
\caption{\label{z_results} Convergence statistics as a function of
  redshift . Top panel: variance (divided by $10^{-2}$). Bottom
  panel: skewness (divided by $10^2$).The solid lines are the
  numerical results. Each line represents a different value obtained after
  smoothing over a different angular scale $\theta$ regularly spaced every
arcminute from 1 to 9 arcminutes. The dotted line shows the
theoretical trend predicted by Bernardeu et al. 1997. The
  numerical trend only weakly depends on the smoothing scale
  $\theta$. The numerical and theoretical results compare fairly well
  at $z>1$, but at $z<1$ the numerical
  results predict a logarithmic slope about two times steeper
than that of analytical theory.}
\end{figure}

\begin{figure}
\begin{center}
\includegraphics[scale=0.39]{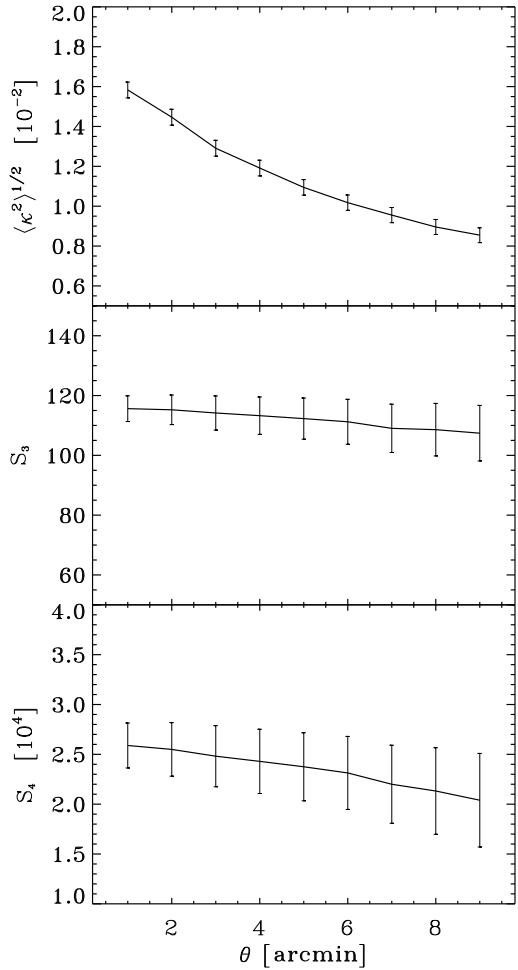}
\end{center}
\caption{\label{theta_results} Convergence statistics as a function
  of smoothing scale. Sources are all located in a single plane at $z=1$.
  Top panel: variance (divided by
  $10^{-2}$). Middle panel: skewness (divided by $10^2$). Bottom
  panel: kurtosis (divided by $10^{4}$). Error bars indicate
  the 1-$\sigma$ dispersion measured around the mean for the 25 cone realisations.}
\end{figure}

In this section, we focus on the measurements of the moments of convergence
($\kappa$) to present the results of the ray-tracing module of
\lemomaf{}.

The dark matter cones we use to make these measurements are $1.4^{\circ}$ on a side, and have
a depth of $4400$ $h^{-1}$ Mpc, with a distance of
$100$ $h^{-1}$ Mpc between lensing planes. In each of these planes, we use a grid with
size $L_{side} = 200$ $h^{-1}$ Mpc which is split in $N_{g} = 800$ cells.
Weak lensing properties and galaxy counts
are measured only in a central field of $1.0^{\circ}$ on a side.
We characterize $\kappa$ through moments of its distribution function
such as its variance $\langle\kappa_{\theta}^2\rangle$, its skewness

\begin{equation}
S_{3}(\theta)=\frac{\langle\kappa_{\theta}^3\rangle}{\langle\kappa_{\theta}^2\rangle^{2}}
\end{equation}
and kurtosis
\begin{equation}
S_{4}(\theta) = \frac{\langle \kappa_{\theta}^{4}\rangle -
  3\langle\kappa_{\theta}^2\rangle^{2}}{\langle\kappa_{\theta}^{2}\rangle^{3}}
\end{equation}
We calculate these moments after smoothing the convergence maps with a
circular top-hat filter of angular scale $\theta$.

We are especially interested in two results to validate our ray
tracing code:\\
1) The dependence of $\langle\kappa_{\theta}^2\rangle^{1/2}$ and
$S_{3}$ as function of the redshift of the source, for which
analytic expressions can be derived.\\
2) The dependence of $\langle\kappa_{\theta}^2\rangle^{1/2}$, $S_{3}$
and $S_{4}$ on the smoothing scale $\theta$, for a given redshift of
the source, for which published values exist in the literature.

In Fig.\ref{z_results} we show our measurements for
$\langle\kappa^{2}\rangle^{1/2}$ and $S_{3}$ as a function of
redshift, for nine different smoothing angles $\theta$ spaced every arcminute from
$1^{\prime}$ to $9^{\prime}$. From this figure we can see that our
computational values follow fairly well the expected theoretical trend.
The agreement is better when the source plane is located at redshifts larger than $z=1$.
For closer source planes, located at $z<1$, the slope of our numerical relation is
 steeper by about a factor of two.

In Fig. \ref{theta_results} we plot our results for
$\langle\kappa^{2}\rangle^{1/2}$, $S_{3}$ and $S_{4}$ as a function of
the smoothing angle $\theta$, for sources located in a single plane at
$z=1$ and a field of view of $1^{\circ}\times 1^{\circ}$. The order of
magnitude of $S_3$ and $S_4$  is consistent with a compilation of
results for these moments made by \cite{vw} for smoothing scales of
$4^{\prime}$. These authors quote an average of $S_{3}\sim 135\pm 10$ and
$S_{4}\sim (3.5\pm 0.5)\ 10^4$ for measures obtained using different methods,
with error bars reflecting the dispersion amongst reported values.
For the same smoothing scale we find $S_3 = 115\pm 5$, and
$S_4=(2.5 \pm 0.3)\ 10^4$, keeping in mind that uncertainties coming from the
numerical simulations themselves can contribute for an extra $10\%$ error in our case,
and that theoretical methods compiled in \cite{vw} suffer from a similar
underestimate.

Being reasonably confident that we are able to produce reliable weak
lensing measurements, we can now couple our ray tracing
code to the outputs of a hybrid model of galaxy formation
to study the source-lens clustering effect on the
convergence statistics, and the effect of weak lensing on galaxy
counts.

\subsection{Source Lens Clustering}

\begin{figure*}
\includegraphics[scale=0.55]{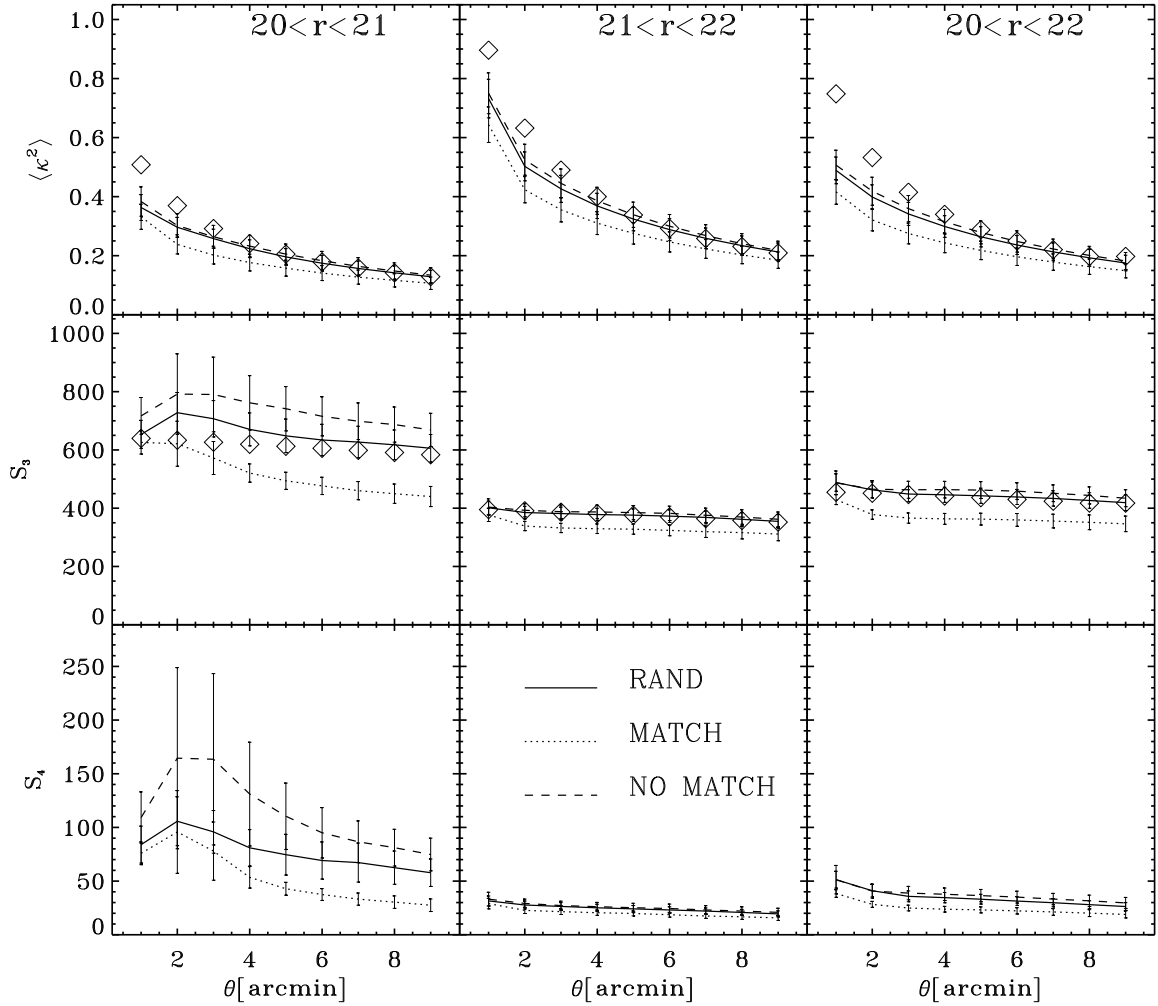}
\caption{\label{pdf_compare_r}
Comparison of the convergence statistics for three
different magnitude ranges in the {\tt SDSS r} filter, and for
three different methods of measuring the convergence at each
galaxy position.
The solid line indicates results obtained with the {\tt RAND} method, the dashed line with the {\tt NO
  MATCH} method, and the dotted line with the {\tt MATCH} method.
$\langle\kappa^{2}\rangle$ has been divided by $10^{-4}$,
$S_{4}$ by $10^4$. Overplotted diamonds show values computed with
the semi-analytic model described in Van Waerbeke et al. 2001.}
\end{figure*}

\begin{figure*}
\includegraphics[scale=0.55]{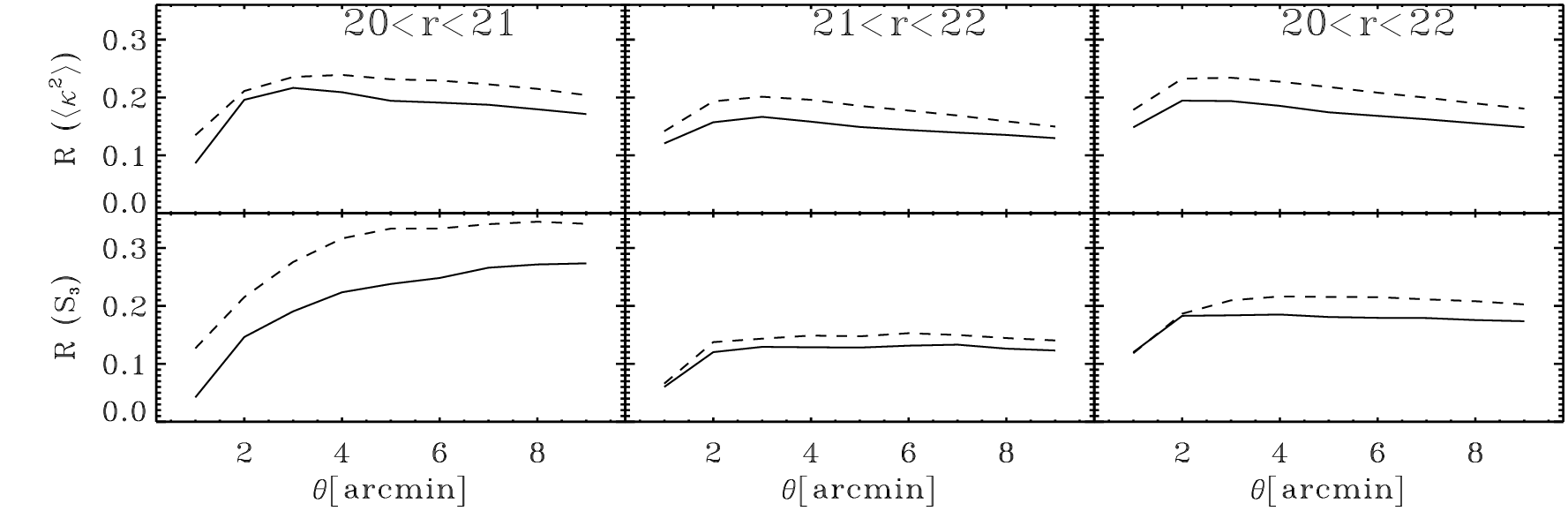}
\caption{\label{slc_percent_r} The $R(\mathcal{S})$ factor as defined in
  Eq. \ref{r_slc} for $\avg{\kappa^2}$ and $S_3$, for the three
  different magnitude ranges in the {\tt SDSS r} filter. The solid
  line compares the methods {\tt MATCH} and {\tt RAND}. The dashed
  line compares the methods {\tt MATCH} and {\tt NO MATCH}.}
\end{figure*}

\begin{figure*}
\includegraphics[scale=0.55]{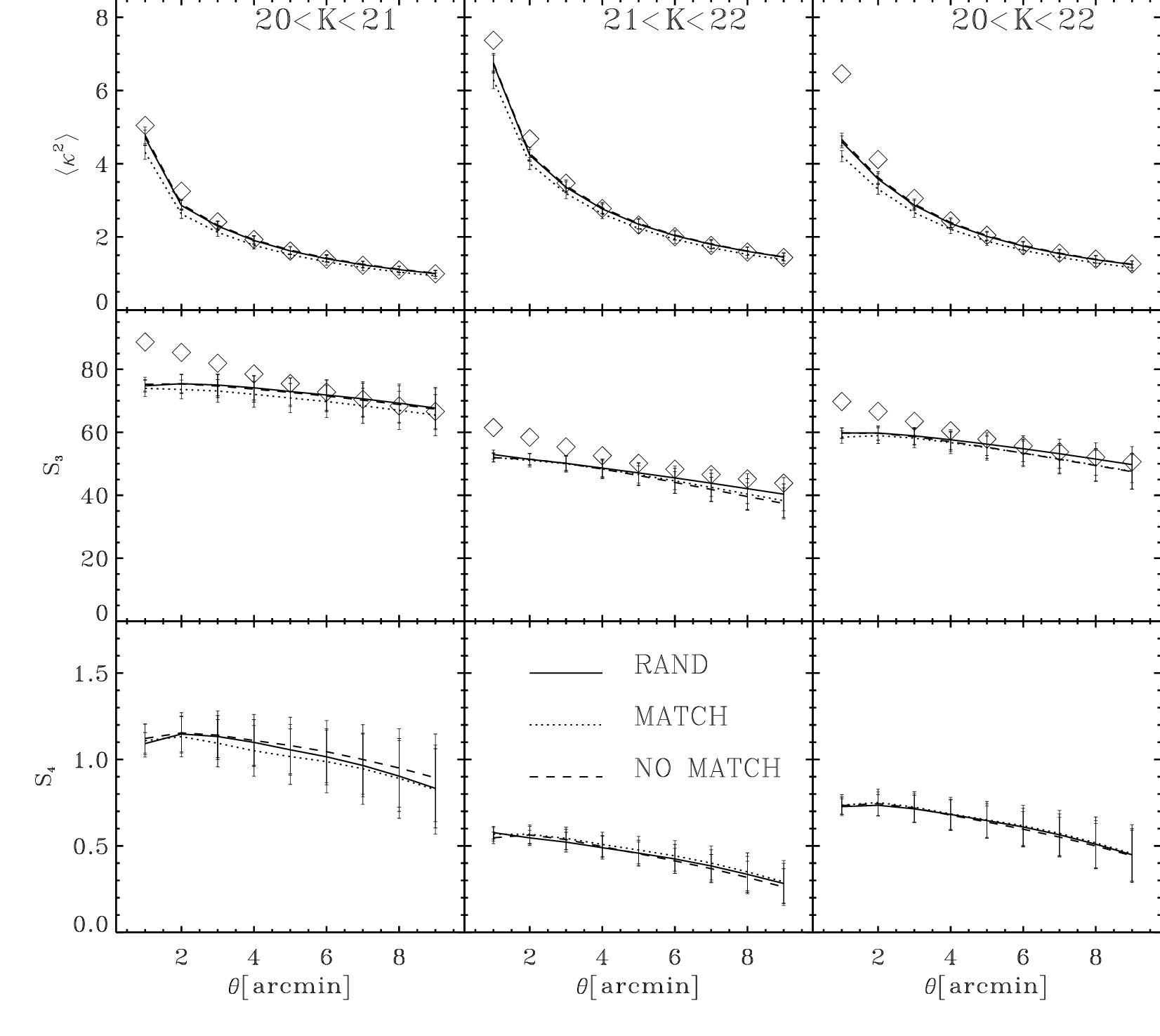}
\caption{\label{pdf_compare_k}Comparison of the convergence statistics for three
  different magnitude ranges in the {\tt JOHNSON K} filter, and for
  three different methods of measuring the convergence at each
  galaxy position.
  The solid line  indicates results obtained with the {\tt RAND} method, the dashed line with the {\tt NO
    MATCH} method, and the dotted line with the {\tt MATCH}
  method. $\langle\kappa^{2}\rangle$ has been divided by $10^{-4}$,
  $S_{4}$ by $10^4$.Overplotted diamonds show values computed with the semi-analytic model described in Van Waerbeke et al. 2001.}
\end{figure*}

\begin{figure*}
\includegraphics[scale=0.55]{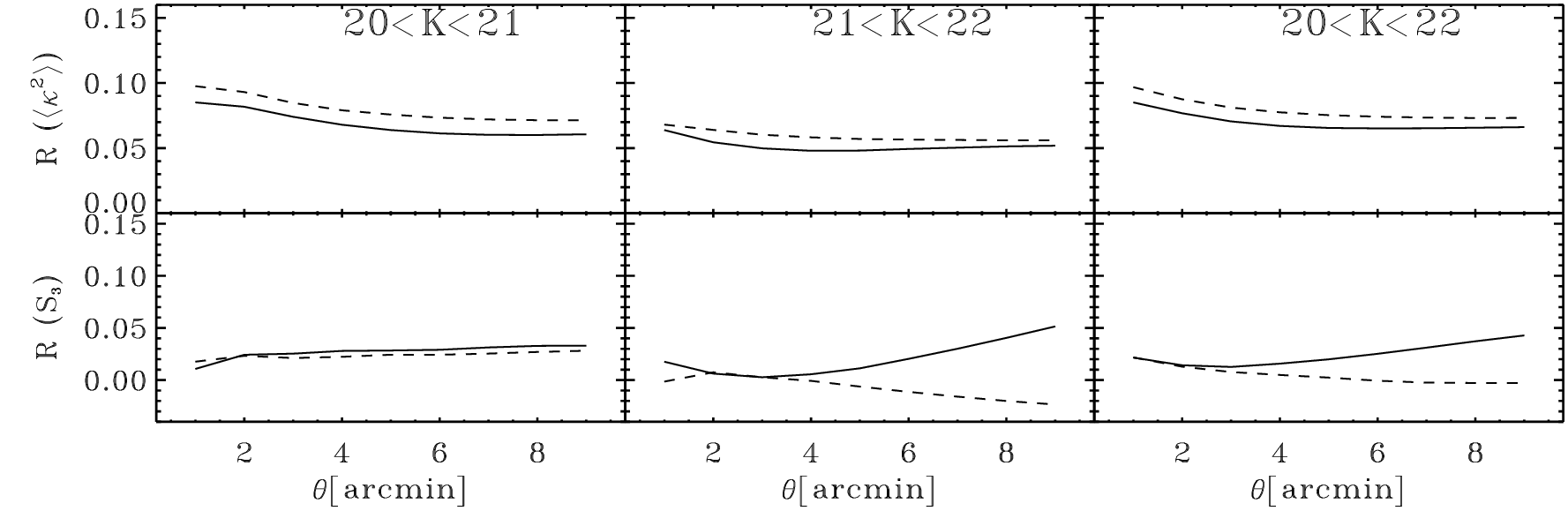}
\caption{\label{slc_percent_k} The $R(\mathcal{S})$ factor as defined in
  Eq. \ref{r_slc} for $\avg{\kappa^2}$ and $S_3$, for the three
  different magnitude ranges in the {\tt JOHNSON K} filter. The solid
  line compares the methods {\tt MATCH} and {\tt RAND}. The dashed
  line compares the methods {\tt MATCH} and {\tt NO MATCH}.}
\end{figure*}

In the previous section we measured the statistics of the convergence
of rays that uniformly covered a fraction of the sky. In reality, we
only have access to the convergence signal measured for galaxies that have
specific clustering properties and, even more important, are correlated
to the lensing potential. This is known in the literature as the
Source-Lens Clustering (SLC) effect.

Theoretical studies of the SLC effect \citep{Bernardeau98} predict
that it alters the higher order statistics of the convergence. More
precisely, while $\langle\kappa^{2}\rangle^{1/2}$ should remain
unchanged according to theory, $S_3$ should have a lower value. As a
matter of fact, $S_{3}$ is known to be sensitive to $\Omega_{m}$
almost independently of $\sigma_{8}$, and a combined analysis of the
skewness and the variance of the convergence could in principle
provide new constraints on the values of $\Omega_{m}$ and
$\sigma_{8}$. Hence the cosmological interest to accurately quantify
the influence of the SLC effect on the higher order statistics of
$\kappa$.

Previous numerical work on the SLC effect \citep{sourcelens} has
focused on its impact on $S_3$ estimations. This work was carried out using
a simple bias model to paint a population of galaxies on top of the dark matter density field.
To our knowledge, our work is the first attempt to use a galaxy
population self-consistently derived from a N-body simulation
to estimate of the impact of the SLC effect.
More specifically, with \momaf{}, we obtain both a dark
matter distribution and its matching galaxy distribution in a light cone, this latter being
derived by post-processing the dark matter simulation with the \galics{} semi-analytic model.
We then calculate the weak lensing signal over
all the field at all the planes used in the ray tracing simulation with \lemomaf{}, and use
this information to shear the properties of galaxies which are in the cone, while
storing the value of the convergence computed at each galaxy position. In this
section, we therefore return to the analysis of the convergence statistics, but only for the
subset of values measured at each galaxy position by interpolation of
the values at the neighboring rays.

To quantify the impact of using a self-consistent modelling of the galaxy population
on our results, we couple the dark matter and galaxy cones in three different
ways. In the first way, that we call {\tt MATCH}, we shear the galaxies
according to their true underlying dark matter distribution. In the second way,
called {\tt RANDOM}, we keep the same pair of galaxy cone/DM cone as in the {\tt MATCH} case,
but erase some of the spatial clustering information by randomizing the
positions of galaxies over the sky
plane while keeping their distance to the observer constant (i.e. we preserve the galaxy
redshift distribution). Finally, in the third case, called {\tt
NO MATCH}, we use the full spatial clustering information for galaxies, but
 shear them using a different underlying dark matter distribution from the one
with which their properties were derived. The idea behind these ``tricks'' is that
cross-comparisons between the {\tt MATCH}, {\tt NO
  MATCH} and {\tt RANDOM} methods should provide us with a better understanding
of where the impact of the source-lens clustering effect on the convergence statistics
comes from.

We use two broad band luminosities for each galaxy, those measured in the {\tt SDSS r} , and
{\tt   JOHNSON K} filters. We build 25 different light cones in order
to minimize the bias effects induced by the random tiling technique, and
maximize the accuracy of the statistics \citep{momaf}. For each light
cone we output both the galaxies and the dark matter distribution.
To mimic observational effects as best as we can, we select galaxies on which
the lensing signal is to be measured based on their apparent magnitude. The resulting redshift distributions
are shown in Fig. \ref{distros}, where we have arbitrarily split magnitudes in three bins:
 $20 < m <21$, $21 < m < 22$ and $20 < m < 22$.
Once again, we recall that each light cone is $1.4^{\circ}$ on a side, and has a depth of
$4400$ $h^{-1}$ Mpc, with a distance between lensing planes of
$100$ $h^{-1}$ Mpc. In each of these planes we use a grid of size
$L_{side} = 200$ $h^{-1}$ Mpc split into $N_{g} = 800$ cells. Weak lensing
properties and galaxy counts are measured in a centered field of $1.0^{\circ}$ on a side.

Fig. \ref{pdf_compare_r} and \ref{pdf_compare_k} show the results
for the higher order moments of the convergence and the six redshift
distributions plotted in Fig. \ref{distros}. Each linetype
corresponds to a different method to construct the maps; {\tt
RANDOM}, {\tt NO MATCH} and {\tt MATCH}. We also compare our
numerical results to the semi-analytic calculations of \cite{VW01}.
Overall, the semi-analytical predictions of these authors show very good agreement
with the lensing signal measured in our mock catalogues. Moreover,
the better agreement with the {\tt RANDOM} mocks is somewhat
expected: out of the three methods we presented here, this is the
one which follows the most closely the assumptions made in
\cite{VW01}.
Most of the small scale deviations from the semi-analytic trend in the variance plots can be attributed to the
finite spatial resolution of our simulation (see Section 3.5).

We introduce the parameter $R$ as done by \cite{sourcelens} to quantify
the amplitude of the SLC effects:
\begin{equation}
R(\mathcal{S}) =  \frac{\mathcal{S}_{\ {\tt NO\ SLC}} -
  \mathcal{S}_{\ {\tt MATCH}}}{\mathcal{S}_{\ {\tt NO\ SLC}}}
\label{r_slc}
\end{equation}
where $\mathcal{S}$ can be $\avg{\kappa^{2}}$ or $S_{3}$, and {\tt NO SLC} is
one of the methods {\tt RAND} or {\tt NO
  MATCH}. Fig. \ref{slc_percent_r} and \ref{slc_percent_k} show the
results for this expression.

From these figures, it is clear that the results for
$\avg{\kappa^{2}}$, $S_3$ and $S_4$ in both filters, and in
the three different magnitude bins show the same trend: these
statistics are systematically lower for the {\tt MATCH} case than
the for the {\tt NO MATCH} and {\tt RANDOM} cases, with these latter
being remarkably similar. Furthermore, we clearly see that the
effect is more pronounced for the narrower redshift distribution of
sources (i.e. {\tt SDSS r} band). These results, which are obtained
using a fairly realistic galaxy distribution, are in broad agreement
with the simpler approach advocated by
\cite{Bernardeau98} and \cite{sourcelens}. However, we also find a significant
SLC effect on the two point statistics ($R \sim 20\%$ for the {\tt SDSS
  r} distributions, $R \sim 5\%$ for the {\tt JOHNSON K}
distributions), which is not expected from  perturbation theory
alone. We suspect that it arises because our simulation probes the
highly non-linear regime. A comparison of the {\tt MATCH} to the {\tt
  NO MATCH} run --- where the clustering of sources is preserved albeit galaxies
positions are not correlated with that of the underlying dark matter distribtion ---
strongly suggests that the signal comes from the adequation of the
intra halo galaxy population with the depth of its host potential
well.

Finally, we would like to emphasize that such a mock catalogue
approach should allow a fast calculation of the SLC effect for
future weak lensing surveys intending to use cosmic shear as a
precision cosmology tool.

\subsection{Galaxy counts}

\begin{figure*}
\begin{center}
\includegraphics[scale=0.40]{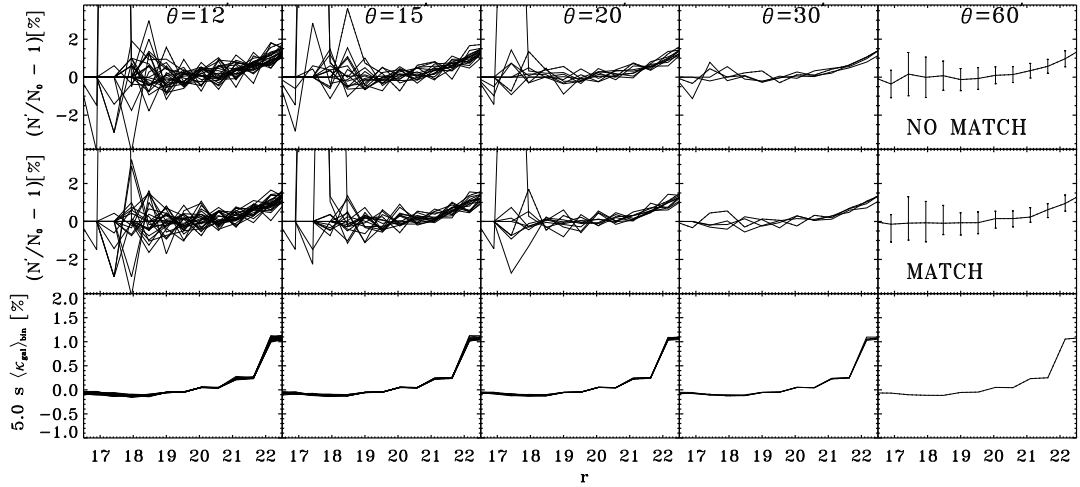}
\end{center}
\caption{\label{counts_compare_r} Percentile difference between
  lensed  ($N^{\prime}$) and unlensed ($N_0$) galaxy counts in the {\tt
    SDSS r} filter, calculated with the {\tt MATCH} (upper panel) and
  {NO MATCH} (middle panel) methods. The lower panel shows the values
  of $\lambda = 5.0\ s\ \avg{\kappa_{gal}}_{bin}$, where $s$ is the logarithmic slope of the
counts and $\avg{\kappa_{gal}}_{bin}$ is the average value of $\kappa$ measured
from the galaxies in that magnitude bin over the field of side size
$1^{\circ}$.Each vertical line splits the plot in panels which contain results
for different patches of angular size $\theta$ on a side, cut inside
an original field of size $\Theta = 60^{\prime}$ on a side. Thus the number of curves
in each row of panels is $N_{p}=(\Theta/\theta) ^2$, and each curve represents the mean of
the measurement over 25 uncorrelated patches. The rightmost upper and middle
panel also show the 1-$\sigma$ dispersion between the 25 $\Theta$ fields ($N_{p} = 1$)
as error bars.}
\end{figure*}

\begin{figure*}
\begin{center}
\includegraphics[scale=0.40]{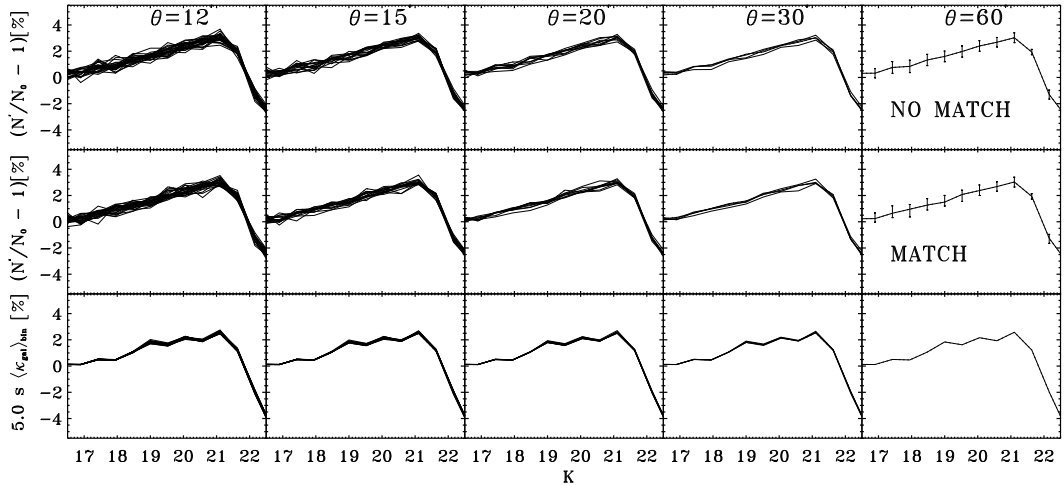}
\end{center}
\caption{\label{counts_compare_k} Same as
  Fig. \ref{counts_compare_r} for the {\tt JOHNSON K} filter. The
  behaviour for magnitudes $K>21$ is a numerical resolution
  artifact which leads to an incompleteness in the number of low luminosity galaxies
 (see also right panel in Fig.\ref{gcounts})}
\end{figure*}

Weak lensing enlarges the area of the sky which is observed,
thus lowering the number density of objects which are detected. In the same time, it causes
galaxies to appear brighter, thus increasing their number density for
a given apparent magnitude. The net effect depends on the slope $s(m)$ of the number
counts of galaxies. Let us write $N_{0}(m) dm$ as the number of
galaxies with magnitudes between $m$ and $m+dm$, and $s(m) = d\log
N_{0}(m)/dm$. If the sources undergo a magnification $\mu = 1/\det\mathbf{A}$
(see section 3.1 for a definition of the matrix A),
and $N^{\prime}(m)$ is the magnified number of sources corresponding to
$N_{0}(m)$, we can write $N^{\prime}(m) = \mu ^{2.5 s - 1} N_{0}(m)$
\citep{Broadhurst, Jain02, Scranton05}.

The effect of the magnification on the solid angle, which is
responsible for a factor $\mu^{-1}$ can only be detected if we use a
grid that allows us to have a resolution in the order of $0.1$
arcmin to make a proper estimation of the deflection angle for a
given galaxy. The angular resolution of our simulations is roughly
$1$ arcmin, which is clearly not enough. Therefore we expect that
only the lensing effects on galaxy magnitudes (which does not
require high angular resolution to be seen) will be perceptible.
This simplifies the expression of magnified number of sources to
$N^{\prime}(m) = \mu ^{2.5 s} N_{0}(m)$, which in turn reduces, in
the weak lensing regime ($\mu = 1 + 2\kappa$), to:

\begin{equation}
  \frac{N^{\prime}(m)}{N_{0}(m)} - 1 = 5.0\  \kappa\  s
  \label{master}
\end{equation}

The numerical experiments we carry out with \lemomaf{} are meant to
explore the effect of weak lensing on galaxy counts via the estimate
of this ratio, as we vary the size of the field in which the measurement of the counts
are performed.

In the 25 original fields of $1^{\circ}\times 1^{\circ}$ we compute
galaxy counts both for unlensed and lensed fields (using the {\tt
  MATCH} and {\tt NOMATCH} methods). We then cut each of these fields into smaller
square patches of angular size $12^{\prime}$, $15^{\prime}$, $20^{\prime}$,
$30^{\prime}$ and $1^{\circ}$ on a side, and measure the counts in all of these sub-fields.
We also calculate for each patch the ratio $N(m)/N_{0}(m)$, and organize our
results as follows. For a patch of size $\theta$, taken
from a field of original size $\Theta$ we have $N_p=(\theta/\Theta)^2$
patches. Labeling subfields as $\ell= 1, \ldots, N_p$
in every original uncut $\Theta$ field, we then proceed to stack together the 25 patches
which have the same index $\ell$. For each patch in the stack, we estimate the ratio of
lensed to unlensed integrated counts, and compute the mean for each
stack.

In Fig. \ref{counts_compare_r} and Fig. \ref{counts_compare_k}
we plot these means for our $N_p$ stacks, for each value of
$\theta$ (column panels) and for both methods, {\tt  NO MATCH} and {\tt MATCH} (top and middle row
panels respectively). In the bottom row of these figures,
we show $\lambda = 5.0\ s\ \kappa$, in order to facilitate the
interpretation of the behaviour of the lensed to unlensed counts ratio
in terms of Eq. \ref{master}. In this intent, we estimate $\kappa$ using an average
value measured for the galaxies which lie in the magnitude bin of interest in the uncut field of
size $\Theta$. From our ray tracing results (see Fig. \ref{z_results}) we know that we
are in the weak lensing regime where $\kappa \ll 1$, so
Eq. \ref{master} tells us that we can expect a depletion in galaxy counts only
when $\avg{\kappa}$ and $s$ have different signs.

In the {\tt SDSS r} band (Fig. \ref{counts_compare_r}),
$\lambda$ (in percent) is restricted to the interval $[0.0, 1.0]$. High
fluctuations in the counts ratio for the low values of the magnitude
are due to an ever smaller number of bright galaxies being present in a subfield
when the subfield size decreases or the brightness of the source increases.
In the magnitude interval $m_{r}=[19, 22]$ where this
effect becomes negligible, we see that galaxy number counts can be enhanced
by up to $1\%$.

In the {\tt JOHNSON K} filter (Fig. \ref{counts_compare_k}), $\lambda$ (in percent)
takes values in the interval $[-4.0,4.0]$, although its change of sign between the
magnitudes $21$ and $23$ is entirely due  to the mass resolution of our N-body simulation
which translates into an incompleteness for galaxies fainter than
$m_{K}=21$ (see also Fig. \ref{gcounts}).
We therefore restrict ourselves to the magnitude interval
$m_{K}=[16,21]$ for which there exists a good agreement between modeled and observed
counts in terms of slope, and the number of bright galaxies per subfield is high
enough. In this magnitude interval, the counts can be enhanced up to a  $3\%$,
for high values of $m_{K}$.

In both filters we note that faint galaxies
are always more enhanced than bright ones: $\lambda$ as a function
of a magnitude is a monotonically increasing function. Of course, this trend
has to break down at some point, when the slope of the faint galaxy counts turns over,
as shown in Fig. \ref{counts_compare_k}, even if in this case it is purely an artifact
due to finite mass resolution in our N-body simulation.

Unsurprisingly, the dispersion
around the theoretically expected ratio increases when the angular size of the field decreases.
Finally, Fig. \ref{counts_compare_r} and \ref{counts_compare_k} show that
the source lens clustering effect does not play an significant role in enhancing
galaxy number counts: the {\tt NO MATCH} and {\tt MATCH} methods pretty much
yield the same quantitative results.

However, from this experiment we confirm our suspicion that the
modification of the solid angle is not resolved in our simulations,
consequently higher angular resolution must be attained if one hopes to
use \lemomaf\  in the study of angular correlations induced by
cosmic magnification.

\section{Discussion and concluding remarks}\label{sec:conclusion}

In this paper we presented the {\it Lensed Mock Map Facility} that combines
dark matter N-body simulations and an hybrid model of hierarchical
galaxy formation to make mock lensed images and convergence maps,
thanks to a ray tracing algorithm.
More specifically, the results presented here were obtained
using a cosmological N-body simulation performed with a treecode, the
\galics{} model of galaxy formation, the \momaf{} pipeline
which constructs galaxy and dark matter light cones, and a
ray tracing algorithm through multiple planes to account
for the weak lensing effect.
This tool suffers from all the shortcomings inherent to each
 of these techniques. However, for \galics{} and \momaf{} these limitations have been
carefully identified in a series of papers published over the past four years.
As far as the ray tracing algorithm is concerned, its limitations in terms of the
N-body simulation parameters have also been discussed quite thoroughly in the literature.
To sum up, these shortcomings impose a limitation on the size of
the fields that can be constructed, and the angular resolution that
can be reached in the construction of the lensed images and the
convergence maps.

Working within the proper interval of validity of these methods, we
performed two numerical experiments with \lemomaf{}. The first one
measured the convergence signal induced by the dark matter density
field at galaxy positions in a light cone. Different methodologies
for this measurement were implemented, in the aim of testing the
consequences of the source-clustering effect on the probability
density function of the convergence. We found that the SLC effect
skews the PDF towards lower values of the convergence and that, in
some cases, it makes this PDF look more gaussian than that obtained
without including SLC, as expected from theoretical considerations.
However, even when probing the \emph{same} dark matter distribution,
the precise trend of the SLC effect depends sensitively on the
specific distribution of the galaxies we consider. For instance, we
demonstrated that a narrower redshift distribution of the sources is
more sensitive to the SLC effect. This could be problematic for
future lensing surveys which intend to perform shear measurement in
{\it thin} redshift slices, a technique called {\it tomography}
\cite{H99}. For the JOHNSON K filter, the SLC effect has an impact
at the few percents level ($2-5\%$) on the estimations of
$\sigma_{8}$ from two point statistics. This level of contamination
was neglected in previous analysis \citep{Bernardeau98,sourcelens},
because its amplitude was well below the largest weak lensing
surveys accuracy at that time (VIRMOS: \cite{VW2000}, RCS:
\cite{Hoek02}). However, future -nearly full sky- missions like SNAP
and LSST will have to reach a precision of $10^{-3}$ on the shear
measurement, i.e. roughly $0.1\%$ from on the shear two-points
statistics (\cite{VWcalibration}). At this level of precision, the
SLC will be a major source of systematics, and the only way to
tackle this issue is to have photometric redshifts for each
individual galaxy, sources and lenses. This strengthen the
requirement that future lensing surveys will have to cover a wide
range of the optical spectrum, from U band to near infrared, with
narrow band filters, similar to the COMBO-17 approach
\citep{Heymans04}.

The second numerical experiment measured the lensed
to unlensed galaxy counts ratio. The value of this ratio was obtained for various
angular sizes of observational fields. The general trend of the
results in the simulation can be understood using Eq. \ref{master}, where
only magnitude changes played an important role in enhancing the
counts in our simulations.
We learnt from this experiment that in order to have a realistic
treatment of the magnification effects over the change of galaxy positions
in the sky we need to go up in angular resolution.

Among the ideas that remain to be investigated using \lemomaf{} are those that
take advantage of mock images at multiple wavelengths to identify the
best strategies for measuring the shear, as well as those
which intend to study the bias introduced by intrinsic alignments or
other systematic effects on this measurement.
Future prospects with \lemomaf{} include the simulation of
galaxy-galaxy lensing and cosmic magnification. This kind of signal
demands a resolution in our simulations in the order of
$0.1$ arcmin. With the N-body simulation we used in this paper,
and FFT methods we barely achieve a resolution of $\sim\ 1 - 2$ arcmin. In
order to reach higher resolutions we do not plan to only rely on more
resolved N-body simulations but also to switch ray tracing strategy
and look in the direction of smooth particle lensing \citep{Aubert},
as we have gained confidence from recent studies with \galics{}
\citep{galicsV} that the small scale distribution of galaxies produced
from an adequate resolution N-body simulation and a new positioning scheme
of galaxies inside the halos can be accurate enough to attain this goal.
We also plan to use \lemomaf{} to help design future lensing surveys, which
will need to employ a tool taylored to tackle non-linear effects such as SLC.

\section*{Acknowledgments}
JEFR thanks the Ecole Normale Sup\'erieure for financial support at
the time of writing the code \lemomaf{} through its ``Selection
Internationale'' project, and Karim Benabed for his much appreciated
guidance during the early stages of this project. We thank Brice
M\'enard for his comments on some of the technical aspects presented
in this paper. The N-body simulation used in this work was run on
the Cray T3E at the IDRIS super-computing facility. This work was
performed in the framework of the HORIZON project.

\appendix
\section{Dark matter} \label{sec:dm}
The cosmological N-body simulation \citep{Ninin99} we use throughout
this paper assumes a flat Cold Dark Matter cosmology with a
cosmological constant ($\Omega_m=1/3$, $\Omega_\Lambda = 2/3$), and a
Hubble parameter $h=H_0/[100$ km s$^{-1}$ Mpc$^{-1}] =0.667$. The
initial power spectrum was taken to be a scale-free ($n_s = 1$) power
spectrum evolved as predicted by \citet{BardeenEtal86} and normalised
to the present-day abundance of rich clusters with $\sigma_8 = 0.88$
\citep{EkeColeFrenk96}. The simulated volume is a cubic box of side
$L_b = 100 h^{-1}$Mpc, which contains $256^3$ particles, resulting in
a particle mass $m_p = 8.272\times 10^9 M_\odot$ and a smoothing
length of 29.29 kpc. The density field was evolved from $z=35.59$ to
present day, and we out-putted about 100 snapshots spaced
logarithmically with the expansion factor.

In each snapshot, we identify halos using a friend-of-friend (FOF)
algorithm \citep{DavisEtal85} with a linking length parameter $b=0.2$,
only keeping groups with more than 20 particles. At this point, we
define the mass $M_{FOF}$ of the group as the sum of the masses of the
linked particles, and the radius $R_{FOF}$ as the maximum distance of
a constituent particle to the centre of mass of the group. We then fit
a tri-axial ellipsoid to each halo, and check that the virial theorem
is satisfied within this ellipsoid. If not, we decrement its volume
until we reach an inner virialised region. From the volume of this
largest ellipsoidal virialised region, we define the virial radius
$R_{vir}$ and mass $M_{vir}$ . These virial quantities are the ones we
use to compute the cooling of the hot baryonic component.
Once all the halos are identified and characterised, we build their
merger history trees following all the constituent particles from
snapshot to snapshot.

\section{Lighting up halos} \label{sec:baryons}
The fate of baryons within the halo merger trees found above is
decided according to a series of prescriptions which are either
theoretically or phenomenologically motivated. The guideline -- which
is similar to other SAMs -- is the following. Gas is shock-heated to
the virial temperature when captured in a halo's potential well. It
can then radiatively cool onto a rotationally supported disc, at the
centre of the halo. Cold gas is turned into stars at a rate which
depends on the dynamical properties of the disc. Stars then evolve,
releasing both metals and energy into the interstellar medium (ISM),
and in some cases blowing part of the ISM away back into the halo's
hot phase. When haloes merge, the galaxies they harbour are gathered
into the same potential well, and they may in turn merge together,
either due to fortuitous collisions or to dynamical friction. When two
galaxies merge, a ``new'' galaxy is formed, the morphological and
dynamical properties of which depend on those of its
progenitors. Typically, a merger between equal mass galaxies will give
birth to an ellipsoidal galaxy, whereas a merger of a massive galaxy
with a small galaxy will mainly contribute to developing the massive
galaxy's bulge component. The Hubble sequence then naturally appears
as the result of the interplay between cooling -- which develops discs
-- and merging and disc gravitational instabilities -- which develop
bulges.

Keeping track of the stellar content of each galaxy, as a function of
age and metalicity, and knowing the galaxy's gas content and chemical
composition, one can compute the (possibly extincted) spectral energy
distribution (SED) of each galaxy. To this end, we use the {\sc
stardust} model \citep{DevriendtGuiderdoniSadat99} which predicts the
SED of an obscured stellar population from the UV to the sub-mm.

\end{document}